\documentclass[preprint,11pt]{elsarticle}
\usepackage{enumitem}
\usepackage{textcomp}

\usepackage{geometry}
\usepackage{mathtools}
\usepackage{color}

\usepackage{amssymb,amsmath}
\usepackage{amsfonts}
\usepackage{graphics}
\usepackage{amsthm}

\newtheorem{remark}{Remark}
\newtheorem{theorem}{Theorem}

\newtheorem{prop}[theorem]{Property}

\def\ga{\gamma}

\def\De{\Delta}

\def\la{\lambda}

\def\N{\mathbb N}

\def\R{\mathbb R}

\def\({\left(}
\def\){\right)}
\def\<{\langle}
\def\>{\rangle}

\DeclareMathOperator{\atan}{atan}

\newcommand\abs[1]{\left|#1\right|}

\begin{document}

\title{Hidden chaotic attractors and chaos suppression in an impulsive discrete economical supply and demand dynamical system}
\vspace{5mm}

\author [rm1,rm2]{Marius-F. Danca\corref{cor1}}
\author [rm3,rm4]{Michal Fe\v{c}kan}

\cortext[cor1]{Corresponding author}
\address[rm1]{Department of Mathematics and Computer Science, Avram Iancu University of Cluj-Napoca, Romania}
\address[rm2]{Romanian Institute of Science and Tecçhnology, Cluj-Napoca, Romania}
\address[rm3]{Department of Mathematical Analysis and Numerical Mathematics, \\
Faculty of Mathematics, Physics and Informatics,\\
Comenius University in Bratislava, Bratislava, Slovak Republic,}
\address[rm4]{Mathematical Institute, Slovak Academy of Sciences, \\
Bratislava, Slovak Republic}

\begin{abstract}Impulsive control is used to suppress the chaotic behavior in an one-dimensional discrete supply and demand dynamical system. By
perturbing periodically the state variable with constant
impulses, the chaos can be suppressed. It is proved analytically that the obtained orbits are bounded and periodic. Moreover, it is shown for the first time that the difference equations with impulses, used to control the chaos, can generate hidden chaotic attractors. To the best of the authors' knowledge, this interesting feature has not yet been discussed. The impulsive algorithm can be used to stabilize chaos in other classes of discrete dynamical systems.
\end{abstract}


\maketitle

\section{Introduction}

Many continuous chaotic systems -such as the law of supply and demand (SD)- containing complicated motions and endogenous cycles endlessly repeating and modeling complex economic realities, have been analyzed in the previous decades.
The discrete systems have broad applications in economics (see e.g. \cite{eco}) and economical SD models, are among the oldest and simplest economic discrete models. As the name implies, a SD model is the interaction between market demand (consumers) and market supply (producers) for a given good or service. It is a classical and effective method of modeling how free market forces establishes the price for a good or service through the quantity supplied and the quantity demanded at any given time (see e.g. \cite{supli1,supli2,supli3} for more information on SD models).

Last years, there came to light the so called ``hidden attractors'', which are difficult to find analytically and also numerically. So, there are no numerical procedures to find trajectories or obits for the passage from equilibria to hidden attractors because the neighborhoods of equilibria do not belong to such attractors.
Therefore, an attractor is called ``self-excited attractor'' if its basin of attraction intersects with any open neighborhood of a stationary state (an equilibrium), otherwise it
is called a ``hidden attractor'' \cite{hid1,hid2,hid3}.
Hidden attractors are important in applications because they might allow unexpected and potentially disastrous responses to perturbations in
a structure like a bridge or aircraft wing. However, on our best knowledge, except the cases of some examples of theoretical models (see e.g. \cite{jaf,jian1,jian2}), there are no investigations on hidden attractors in real applied examples of chaotic maps, such as the considered SD model. Thus, the sudden appearance of some hidden chaotic attractor, could represents an important disadvantage for the system. The consequences could be dramatic such as in the pilot-induced oscillations that entailed the YF-22 crash in April 1992 and Gripen crash in August 1993 \cite{nini}. Related to this accident, in \cite{nini2} one can read: ``Since stability in simulations does not imply stability of the physical control system (an example is the crash of the YF22), stronger theoretical understanding is required''. Therefore, in the simulations of the stabilization or control of chaos, identifying hidden attractors should be a mandatory task.
It is understandable that avoiding chaos in economical systems, especially the hidden chaotic behavior, which is much more difficult to identify and control, is a desirable phenomena. Therefore, many researchers suggested applying chaos control in financial systems in order to improve their performances such as preserving stability. Indeed, controlling a chaotic market model may lead to economic efficiency.

On the other side, changes in the system variables, in the form of instantaneous pulses (method studied by G\"{u}uemez and Mat\`{\i}as \cite{gm1,gm2})), proved to be of a real interest in chaos control.

The existence of hidden attractors, in the case of the SD model, might be related to a multistability-like, i.e. starting from an assigned initial condition, the system remains on one or other attractor, depending on external perturbations \cite{multi1,multi2,multi3,multi4}.

The impulsive algorithm, presented here, is useful when the system parameters are inaccessible, namely in the cases of certain chemical experiments, economic policy, biological and ecological systems, electrical circuits etc.

In this paper we show that this method can be used to suppress the chaos in the classical law of SD model.

Let the following discrete dynamical system

\begin{equation*}\label{unu}
x_{n+1}=f(x_n),~~~
\end{equation*}
\noindent with $f$ some continuous map of $\mathbb{R}$ into itself, and the orbit $x_n$ with $n\in \mathbb{N}$, starting from an initial condition $x_1\in \mathbb{R}$. Here $\mathbb{R}=(-\infty,\infty)$ and $\mathbb{N}=\{1,2,\cdots\}$.

Consider that the system behaves chaotically. If every $\Delta$ steps one perturbs the state variable $x$ with a positive or negative preferable small quantity, $\ga$: $x\leftarrow x+\gamma$, it is possible to stabilize the system chaotic behavior.

The impulsive control can be modeled as the following impulsive difference equation (IDE)
\begin{equation}\label{e1d}
x_{n+1}=\left\{
\begin{array}
[c]{l}%
f(x_n),~n\in \N,\\
x_n+\gamma,~\text{if~~} n=n_{i},%
\end{array}
\right.
\end{equation}
where $\{{n_i}\}_{i\in \mathbb{N}}$, $n_i=i\Delta=\Delta, 2\Delta,...$,  is an increasing sequence of natural numbers.

Compared to existing control algorithms which use parameter changes, and require generally
a supplementary knowledge of some data related to the system as Lyapunov exponents, covariant and contravariant vectors etc, the impulsive
algorithms, which change the system variables, are much easier to implement.

For example applications of impulsive control of discontinuous chaotic systems of fractional order can be found in \cite{ex1}, for integer-order discontinuous systems see \cite{ex2}, the case of continuous fractional-order systems can be found in \cite{ex3,ex4,danca1}, and for discrete-time systems see \cite{dancax,dancay}).

The discrete Zhang SD model, for which the price expectation, as described in \cite{Z} (see also \cite{futo}) is
\begin{equation}\label{supl}
x_{n+1}=f(x_n):=(1-\lambda)x_n+\frac{a}{b}\lambda-\frac{\lambda \atan(\mu x_n)}{b},~~~ n\in \mathbb{N},
\end{equation}

\noindent with $\lambda$, a, b, $\mu$, given parameters.

It is considered that there is a time lag in supply, because producers must decide how much to produce before prices are observed in the
market. Therefore, the quantity supplied will be given as a function of the expected price, and not the actual price, i.e. $x_{n+1}$ as function of $x_n$. Also, note that expected price
simply refers to the mathematical expected value of the price \cite{futo}.

\begin{figure*}
\begin{center}
\includegraphics[scale=0.5] {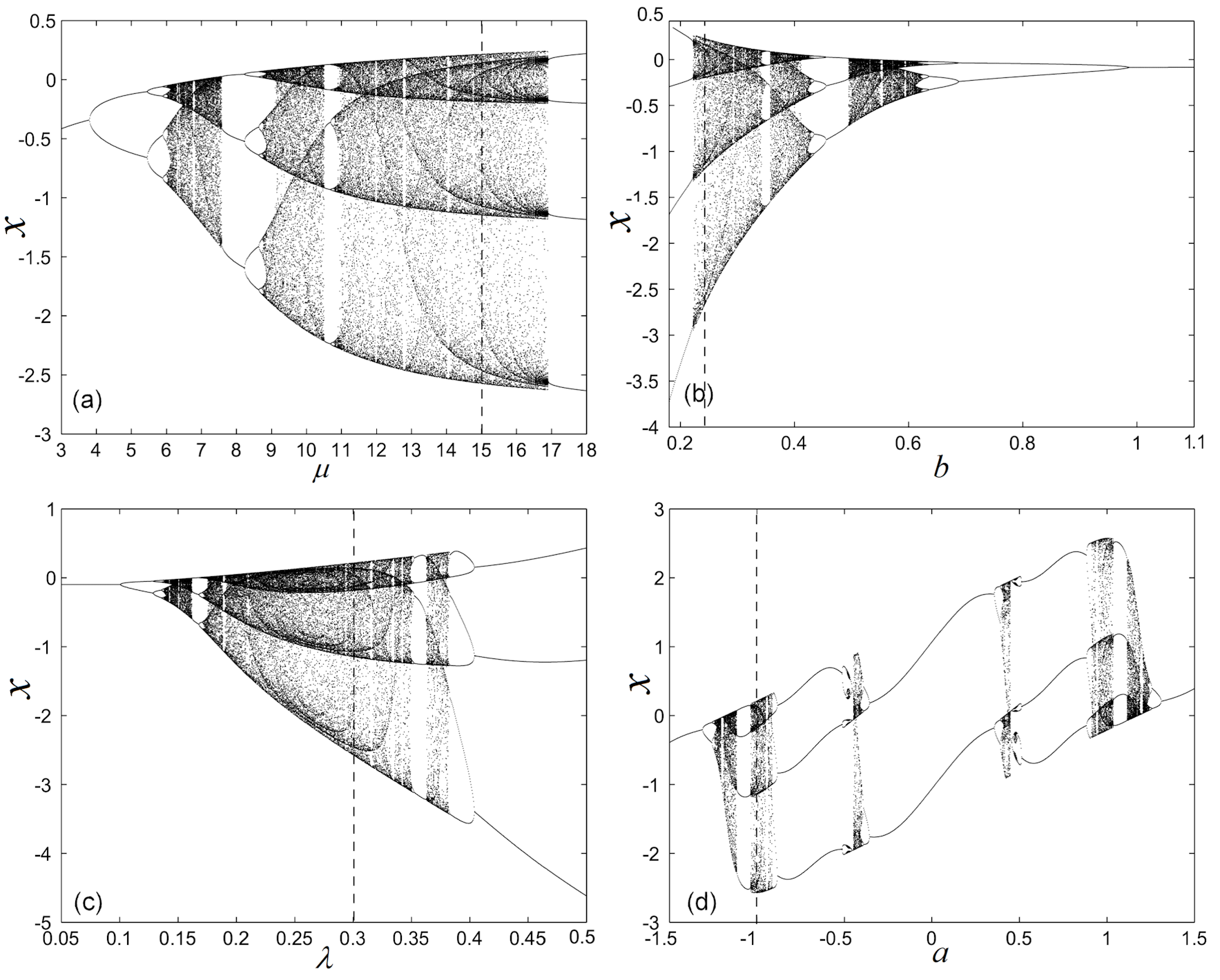}
\caption{Bifurcation diagrams of the not impulsed demand and supply model \eqref{supl}: (a) $\mu\in[3,18]$ and $a=-1,b=0.25$, $\la=0.3$; (b) $b\in[0.2,1.1]$ and $a=-1$, $\mu=15$ and $\la=0.3$; (c) $\la\in[0.05,0.5]$ and $a=-1,b=0.25$, $\mu=15$; (d) $a\in[-1.5,1.5]$ and $b=0.25$, $\mu=15$ and $\la=0.3$.}
\label{fig00}       
\end{center}
\end{figure*}

When monitoring the system parameters of the unperturbed system, period doubling cascade bifurcation, reverse bifurcations, coexisting bifurcations for variations of any of the parameters $\mu$, $\la$, $a$ or $b$ are found (figure \ref{fig00}).

In this paper we are interested in suppressing the chaotic evolution of the price expectation by using impulsive control \eqref{e1d} and also in revealing the existence of the hidden chaotic attractors and self-excited stable cycles.


The paper is structured as follows: in Section 2 an analytical study of the impulsive SD model is presented; Section 3 is dedicated to numerical results on the chaos stabilization and Section 4 presents coexisting attractors and hidden chaotic attractors. The last section presents some conclusions.

\section{Analytical study of the impulsed supply and demand system }

Beside finding $\Delta$ and $\gamma$ necessary to achieve the chaos stabilization, boundedness is another important issue: will obtained impulsed orbits remain bounded while the system is perturbed and, also, there exist periodic stable orbits?

To answer to this question, following \cite{Z} (see also \cite{futo}), we assume that $b >0$, $\mu > 0$ $\lambda\in(0,1)$.

Let $\{n_i\}_{i\in\N},$ $n_i=i\De$ and $\De\in\N$. Consider the IDE \eqref{e1d}-\eqref{supl} in the following equivalent form
\begin{equation}\label{e1}
x_{n+1}=\begin{cases}f(x_n),\quad n\in\N\setminus\{n_i\}_{i\in\N},\\
                      f(x_n)+\gamma,\quad n=n_i,\end{cases}
\end{equation}
for a number $\gamma\in\R$.

Then the dynamics of \eqref{e1} is given by the mapping
\begin{equation}\label{e2}
F(x)=f^{\De}(x)+\ga,
\end{equation}
since
\begin{equation}\label{e2a}
x_n=f^{n-1-\lfloor \frac{n-1}{\De}\rfloor\De}(F^{\lfloor\frac{n-1}{\De}\rfloor}(x_1)),\quad n\in\N,
\end{equation}
where $\lfloor\cdot\rfloor$ is the floor function. In particular, \eqref{e2a} implies
\begin{equation}\label{e2b}
x_{i\De+1}=F^i(x_1),
\end{equation}
for any $i\in\N$. Then \eqref{e2b} gives that any $p$-periodic
orbit of $F$ generates a $p\De$-periodic orbit of \eqref{e1}. Let
$k_-\le x_1\le k_+$ for some $k_-<k_+$. Then
\begin{equation}\label{e3b}
\begin{aligned}
(1-\la)x_{k\De}-A\la+\ga\le x_{k\De+1}&\le (1-\la)x_{k\De}+A\la+\ga\\
(1-\la)x_{i\De+j}-A\la&\le x_{i\De+j+1}\le
(1-\la)x_{i\Delta+j}+A\la
\end{aligned}
\end{equation}
for $j=1,2,\cdots,\De-1$, $k\in\N$ and $i\in\N_0=\N\cup\{0\}$, where
\begin{equation}\label{e2c}
A=\frac{|a|}{b}+\frac{\pi}{2b}.
\end{equation}
So one has
\begin{equation}\label{e3}
\begin{gathered}
(1-\la)^{j-1}x_{i\De+1}-A(1-(1-\la)^{j-1})\le x_{i\De+j}\\
\le (1-\la)^{j-1}x_{i\De+1}+A(1-(1-\la)^{j-1})\quad\textrm{for
} j=1,\cdots,\De,i\in\N_0.
\end{gathered}
\end{equation}
First, we derive from \eqref{e3b} and \eqref{e3}
\begin{equation}\label{e4}
|x_{i\De+j}|\le (1-\la)^{j}|x_{i\De}|+|\ga|+A\quad\textrm{for }
j=1,\cdots,\De,i\in\N.
\end{equation}
Then \eqref{e3b} and \eqref{e4} give
$$
\begin{aligned}
|x_{i\De}|&\le (1-\la)^{(i-1)\De}|x_{\De}|+\left(|\ga|+A\right)\frac{1-(1-\la)^{(i-1)\De}}{1-(1-\la)^{\De}}\\
&\le |x_{1}|+A+\frac{|\ga|+A}{1-(1-\la)^{\De}}\quad\textrm{for }
i\in\N.
\end{aligned}
$$
Then \eqref{e4} implies
\begin{equation}\label{e5}
|x_{n}|\le
|x_{1}|+\frac{|\ga|+A}{1-(1-\la)^{\De}}+|\ga|+2A\quad\textrm{for
} n\in\N.
\end{equation}
Consequently, iterations of \eqref{e1} are bounded.

Next, by $x_1\in[k_-,k_+]$, we obtain from \eqref{e3b} and \eqref{e3}
\begin{equation}\label{e6}
\begin{gathered}
(1-\la)^{\De}k_--A(1-(1-\la)^{\De})+\ga\le F(x_1)\\
\le (1-\la)^{\De}k_++A(1-(1-\la)^{\De})+\ga.
\end{gathered}
\end{equation}
Assuming
\begin{equation}\label{e7}
(k_-+A)(1-(1-\la)^{\De})\le \ga \le
(k_+-A)(1-(1-\la)^{\De}),
\end{equation}
\eqref{e6} implies
$$
k_-\le F(x_1)\le k_+\quad\textrm{for } x_1\in[k_-,k_+].
$$
The Brouwer fixed point theorem gives the existence of a fixed point $x_1^*\in[k_-,k_+]$ of $F$, which implies the existence of $\De$-periodic orbit of \eqref{e1} starting from $x_1^*$ (see \eqref{e2b}). Since $F : [k_-,k_+]\to[k_-,k_+]$ then \eqref{e1} may have much more sophisticated dynamics \cite{El}. Note that \eqref{e7} has a sense if and only if
\begin{equation}\label{e8}
2A\le k_+-k_-.
\end{equation}
Summarizing, we arrive at the following result.

\begin{theorem}\label{th1}
Consider \eqref{e1} with $b>0$ and $\la\in(0,1)$. Then all iterations of \eqref{e1} are bounded on $\N$. If \eqref{e8} holds with \eqref{e2c}, then taking $\ga$ satisfying \eqref{e7}, \eqref{e1} has a $\De$-periodic orbit starting from $[k_-,k_+]$.
\end{theorem}

Furthermore, we derive
\begin{equation}\label{derivata}
f'(x)=1-\la-\frac{\la\mu}{b(\mu^2x^2+1)}.
\end{equation}
Using
$$
F'(x_1)=f'(x_{\De})\cdots f'(x_1),
$$
we obtain
$$
|F'(x_1)|=|f'(x_{\De})\cdots f'(x_1)|\le \left(1-\la+\frac{\la\mu}{b(\mu^2k_0^2+1)}\right)^{\De}
$$
for $k_0=\min\{|k_-|,|k_+|\}$ and any $x_1\in[k_-,k_+]$.

By applying the Banach fixed point theorem, we arrive ar the following result.
\begin{theorem}\label{th2d}
Consider \eqref{e1} with $b>0$ and $\mu>0$ and $\la\in(0,1)$. Assume \eqref{e8} and if
\begin{equation}\label{e12d}
\frac{\mu}{\mu^2k_0^2+1}<b
\end{equation}
for $k_0=\min\{|k_-|,|k_+|\}$ and $\ga$ satisfying \eqref{e7}. Then there is a unique $\De$-periodic orbit of \eqref{e1} in $[k_-,k_+]$ which is in addition exponentially stable with a convergence rate factor $\left(1-\la+\frac{\la\mu}{b(\mu^2k_0^2+1)}\right)^{\De}$.
\end{theorem}
Note \eqref{e12d} holds if and only if
\begin{equation}\label{e12e}
\mu\notin \begin{cases} \left[\frac{1-\sqrt{1-4b^2k_0^2}}{2bk_0^2},\frac{1+\sqrt{1-4b^2k_0^2}}{2bk_0^2}\right]&\quad \text{if}\quad 0<2bk_0\le1,\\
                        \emptyset&\quad \text{if}\quad 2bk_0>1,\\
                        [b,\infty)&\quad \text{if}\quad k_0=0.
\end{cases}
\end{equation}
Hence taking a sufficiently long interval $[k_-,k_+]$, i.e. assuming \eqref{e8}, we can adjust $\gamma\in\R$, i.e. assuming \eqref{e7}, so that iterations of \eqref{e1} starting in  $[k_-,k_+]$ are uniformly bounded (see \eqref{e5}). Then we can adjust $\mu$, i.e. assuming \eqref{e12e}, so that these iterations are attracted by an exponentially stable $\De$-periodic orbit.

\begin{remark}
\itemize
{\item [(i)]Two classical ways for finding fixed points have been used: one is the Brouwer fixed point theorem applied in Theorem 1 and the second one is the Banach fixed point applied in Theorem 2. Moreover, some conditions for parameters to get these fixed points are necessary. In general, it is difficult to find fixed points. As presented in this paper (see also \cite{ei}), for concrete parameters, the fixed points can be found numerically.
\item[(ii)] The above boundedness results are true because of the particular form of the system and also because the impulses are periodic. On the other side, for general systems, small perturbations may lead to unbounded oscillations for particular values of parameters.}
\end{remark}

Summarizing, one concludes that for every $\Delta$ and $\gamma$, the impulsed system \eqref{supl} has bounded orbits and there exist periodic orbits (see \cite{ei} for more details on difference equations with impulses).

\section{Chaos suppression in the supply and demand system}

In this section the impulsed system \eqref{e1d}-\eqref{supl} is considered for several values of $\Delta$. In order to can choose the $\ga$ values for some chosen $\De$, the bifurcation analysis with respect $\ga$ was carried out. Parameters values are chosen as $a=-1,b=0.25$, $\mu=15$ and $\la=0.3$, representing values for which the not impulsed system presents a significant chaotic behavior (see dotted lines in figure \ref{fig00}). As can be seen, period-doubling and reverse bifurcations typically for Feigenbaum route to chaos for the logistic map may appear.

In this paper \emph{numerical chaos} is considered in which the Liapunov exponent is numerically observed to be positive.

For the considered set of parameters, one obtains the following results.

\begin{theorem}\label{teo}
 For $a=-1,b=0.25$, $\mu=15$, $\la=0.3$, and $x_1=0.35$, the impulsed system \eqref{e1d}-\eqref{supl} admits bounded iterations and a unique $\De$-periodic exponentially stable orbit.
\begin{proof}
$A$, given by \eqref{e2c} is

$$
A=\frac{|a|}{b}+\frac{\pi}{2b}=10.28.
$$
Let consider $k_-=-11$ and $k_+=+11$, so that the initial condition $x_1$ verifies the relation $k_-<x_1<k_+$ and also the relation \eqref{e8}, $2A=21.56<k_+-k_-=22$. Then
$$
\frac{\mu}{\mu^2k_0^2+1}=5.51e-4<0.25=b,
$$
where $k_0=11$.
Next, Theorem \ref{th2d} applies.
\end{proof}
\end{theorem}

Note also, that $2bk_0=5.5>1$ and the relation \eqref{e12e} ensures the applicability of relation \eqref{e12d}.

Computationally, the impulsive algorithm acts as follows: every iteration step, the new value of the variable $x_{n+1}$ is determined with the relation \eqref{supl} and, if $n~ (mod~ \Delta)=0$, then $x_{n+1}$ is perturbed as follows: $x_{n+1}=x_{n+1}+\ga$.

The presented images present bifurcation diagrams versus $\gamma$, time series, cobweb diagrams (first return maps), Lyapunov exponent (LE), and attraction basins. The LE are determined with the known relation

\[
LE=\lim_{n\rightarrow \infty}\frac{1}{n}\sum_{i=1}^{n-1} \ln|(f'(x_i)|,
\]
which numerically, via \eqref{derivata}, becomes

\[
LE\approx \frac{1}{n}\sum_{i=1}^{n-1} \ln\abs{\frac{\lambda\mu}{b(\mu^2x_i^2+1)}+\lambda-1}.
\]

Next, the finite-time LE is considered with $n=1000$ which, within the underlying attraction basin, provide with good numerical approximations (three significant digits), the same result.

To get rid of transient behavior, the first transient iterates have been dropped.

Next, few significant cases are considered.

\begin{figure*}
\begin{center}
\includegraphics[scale=0.63] {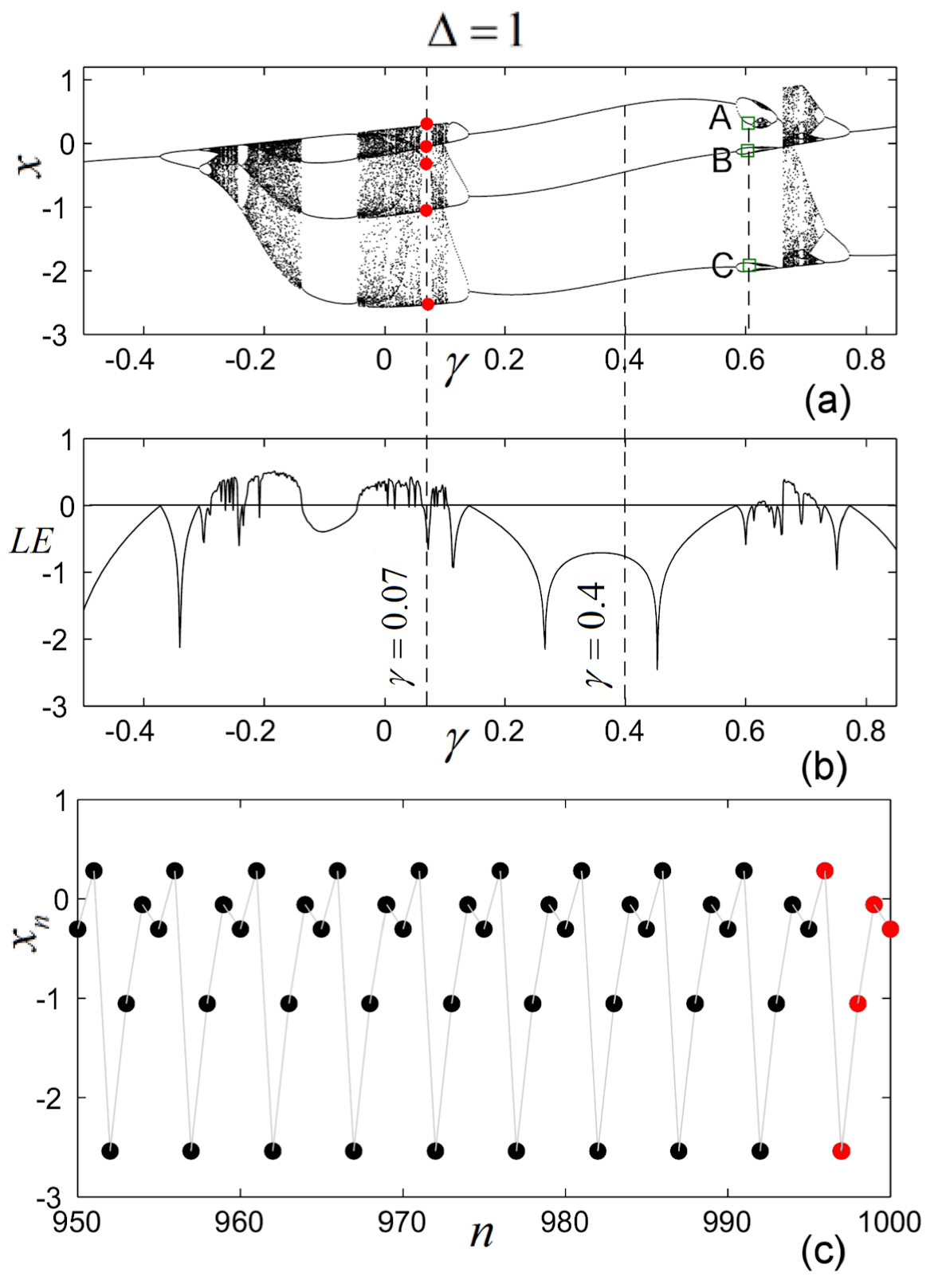}
\caption{Chaos suppression of the impulsed SD model \eqref{e1d}-\eqref{supl} for $\De=1$ and $\ga=0.07$: (a) Bifurcation diagram; (b) Lyapunov exponent; (c) Time series for the last $100$ iterations. The 5 red doted points represents the obtained stable periodic cycle.}
\label{fig1}       
\end{center}
\end{figure*}

\begin{figure*}
\begin{center}
\includegraphics[scale=0.63] {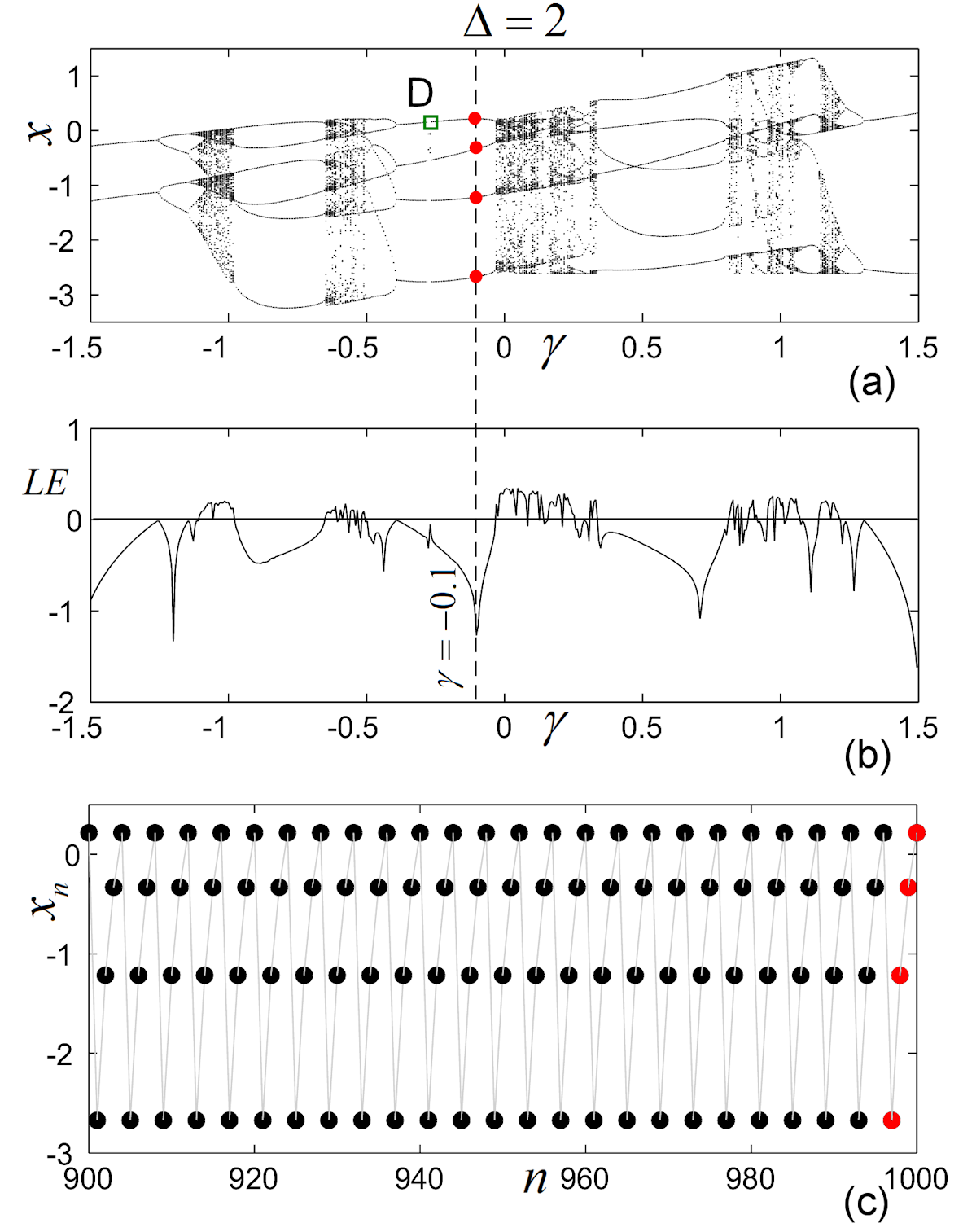}
\caption{Chaos suppression of the impulsed SD model \eqref{e1d}-\eqref{supl} for $\De=2$ and $\ga=-0.1$: (a) Bifurcation diagram. (b) Lyapunov exponent; (c) Time series for the last $100$ iterations. The 4 red dotted points represents the obtained stable periodic cycle.}
\label{fig2}       
\end{center}
\end{figure*}

\begin{figure*}
\begin{center}
\includegraphics[scale=0.6] {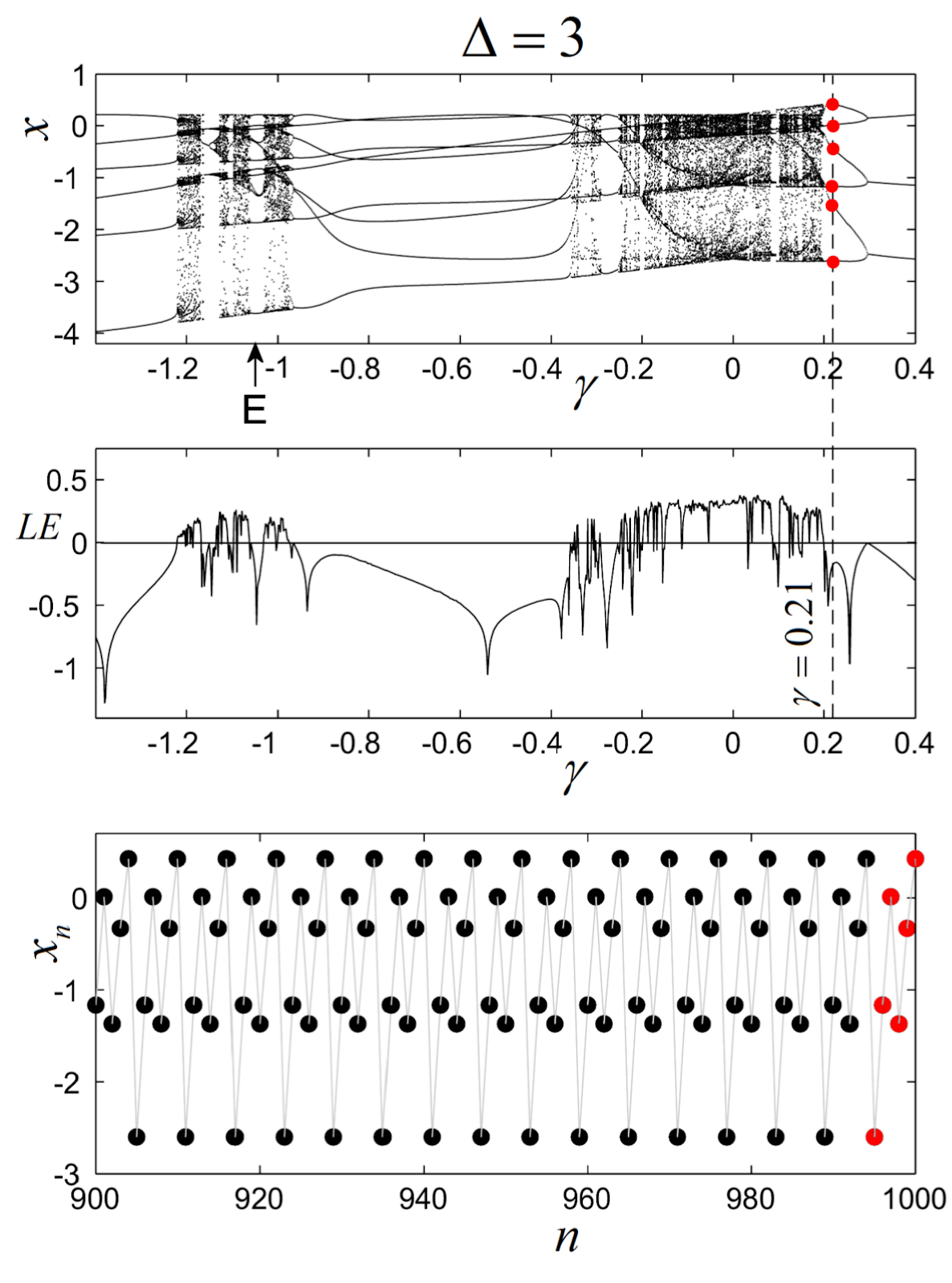}

\caption{Chaos suppression of the impulsed SD model \eqref{e1d}-\eqref{supl} for $\De=3$ and $\ga=0.21$: (a) Bifurcation diagram. (b) Lyapunov exponent; (c) Time series for the last $100$ iterations. The 5 red dotted points represents the obtained stable periodic cycle.}
\label{altii}       
\end{center}
\end{figure*}

\begin{figure*}
\begin{center}
\includegraphics[scale=0.63] {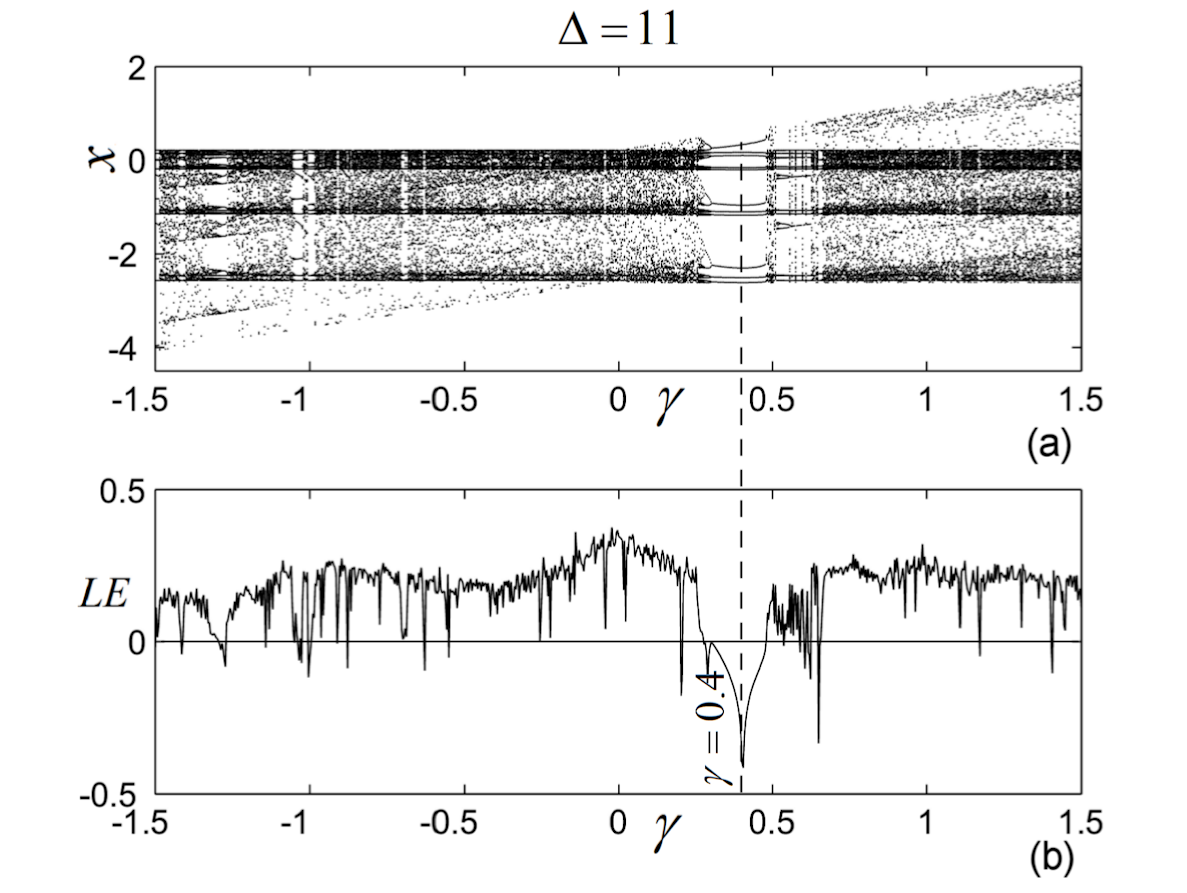}
\caption{Chaos suppression of the impulsed SD system \eqref{e1d}-\eqref{supl} applied every $\De=11$ steps; (a) Bifurcation diagram; (b) LE.}
\label{figx}       
\end{center}
\end{figure*}

\begin{enumerate}[leftmargin=*]
\item $\Delta=1$.
In this case, the SD system is impulsed every iteration. From the bifurcation diagram, and LE versus $\gamma$ (figure \ref{fig1} (a),(b)), one can see there are several periodic windows, containing $\gamma$ values which determine stable orbits in the impulsed system. For example, for $\gamma=0.07$, one obtains the stable 5-period cycle, revealed by the time series in figure \ref{fig1} (c).
Because $\ga$ is positive, one can consider the system receives some external ``energy'' every step.
\item $\Delta=2$. Not only positive $\ga$ values can stabilize the chaos. For example, for $\De=2$, when the system is impulsed every two iteration steps, consider $\ga=-0.1$, chosen within the largest stable periodic window where (see the bifurcation diagram in figure \ref{fig2} (a)). A stable 4-period cycle revealed by the LE and time series (figure \ref{fig2} (b), (c)) is obtained.

\item $\Delta=3$. In this case, to obtain a stable cycle, from the bifurcation diagram (figure \ref{altii} (a)) one can choose $\gamma=0.21$, which generates the stable cycle of period 6 (see the LE in figure \ref{altii} (b) and time series in figure \ref{altii} (c)).
\item Large $\De$ values. Note that even for large values of $\De$, the chaos still can be suppressed (see the bifurcation diagram for $\De=11$ in figure  \ref{figx}, where one can see there are several periodic windows)\footnote{In order to avoid possible influence of previous $\De$ values, $\De$ has been chosen a prime number.}. This property could be useful in practice, because rare interventions (impulses) are preferable to frequently perturbations.
\end{enumerate}

\begin{remark}
\begin{itemize}\
\item[i)]As bifurcation diagrams reveal, for every values of $\De$, impulses applied to the SD system does not change significantly the topological aspects of the attractors of underlying not-impulsed system.
\item[ii)] The impulsive algorithm can be used not only for chaos suppression, but also for chaoticization (anti-control of chaos), by perturbating the state variables periodically or randomly.
    \end{itemize}
\end{remark}

\section{Hidden chaotic attractors in the impulsed supply and demand system}
Multistability or coexistence of different attractors, one of generally necessary ingredients for hidden attractors, and the existence of hidden attractors for some given set of parameters and different initial conditions, are one of the most exciting phenomena in dynamical systems.

As known, the existence of hidden attractors is related to equilibria. So, if the orbits starting from small neighborhoods of unstable equilibria reach an attractor (chaotic or stable cycle), the attractor is called \emph{self-excited}. Otherwise, if the attractor cannot be obtained with orbits starting from small neighborhoods of the unstable equilibria (the basin of attraction of the attractor does not overlap with a small vicinity of unstable equilibria), the attractor is called \emph{hidden }and special numerical procedures have to be developed to find it. Determining initial conditions which lead to hidden attractors (stable cycles or chaotic attractors) is still a challenge for numerical procedures to find hidden attractors. Some steps regarding an analytical-numerical procedure have been done by Kuznetsov and coworkers (see e.g., \cite{hid1}).

Because equilibria of the impulsed SD system are difficult to find analytically, the hidden attractors are determined numerically aided by computer graphics.

\begin{figure*}
\begin{center}
\includegraphics[scale=0.65] {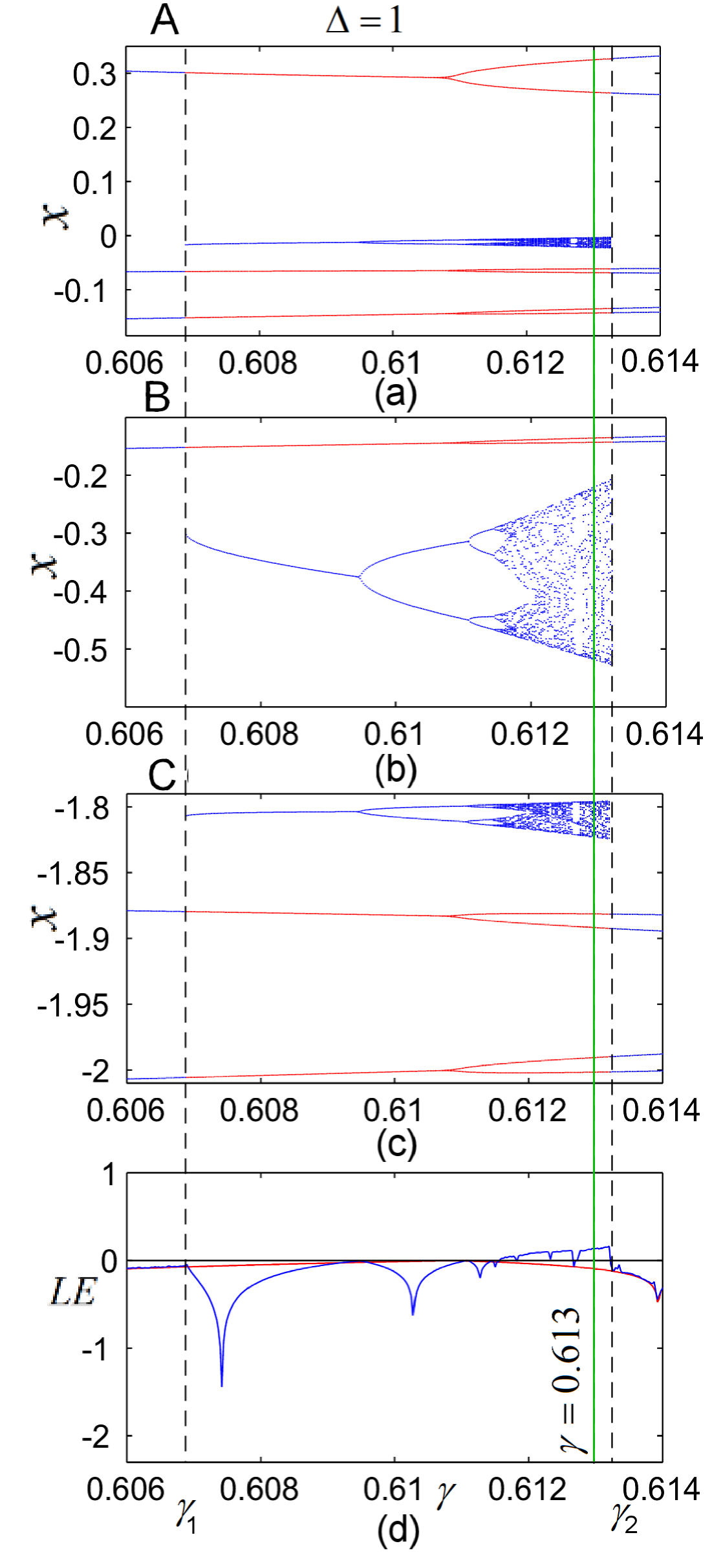}
\caption{Coexisting attractors: stable cycle (red plot) and hidden chaotic attractor (blue plot) in the case $\De=1$ identified for $\ga\in[\ga_1,\ga_2]$, with $\ga_1\approx0.6065$ and $\ga_2\approx0.6135$, obtained from two different initial conditions: (a)-(c) Zoomed details of zones A,B and C from the bifurcation diagram in figure \ref{fig1}; (d) Overplotted LEs. }
\label{fig5}       
\end{center}
\end{figure*}

Let consider the case $\De=1$. Several bistable narrow windows in the $\ga$ space containing coexisting attractors have been found. In figure \ref{fig5} the details A,B and C from the bifurcation diagram in figure \ref{fig1} are zoomed, for $\ga\in[0.606,0.614]$. The two overploted bifurcation diagrams and LE respectively (blue and red plot) are obtained by running the algorithm from different initial conditions. As can be seen in the zoomed details A,B and C, there exist two values $\ga_1\approx 0.6065$ and $\ga_2\approx0.6135$ between which there coexist two different attractors: a stable cycle with negative LE (red plot in figure \ref{fig5} (d)) and a chaotic attractor with positive LE (blue plot in figure \ref{fig5} (d)). To note that the chaotic attractor is born at $\ga_1$ by following the standard bifurcation cascade of the logistic map, and suddenly vanishes at $\ga_2$, while the stable cycle (red plot) persists continuously for $\ga\in[0.606,0.614]$. It seems that the system  presents a memory-like phenomenon of the previously separated regions and, after $\ga_2$, the system suddenly jumps to the previous stable state.

\begin{prop}
Let the impulsed SD system \eqref{e1d}-\eqref{supl}, with $\De=1$. There exists at least one value $\ga\in(0.606,0.614)$, for which the impulsed system admits a hidden chaotic attractor coexisting with a self-excited stable cycle.
\end{prop}

The proof is done numerically aided by computer graphics analysis, by studying the attraction basins and the behaviour of the orbits starting in neighborhood of unstable equilibria.

\begin{figure*}
\begin{center}
\includegraphics[scale=0.65] {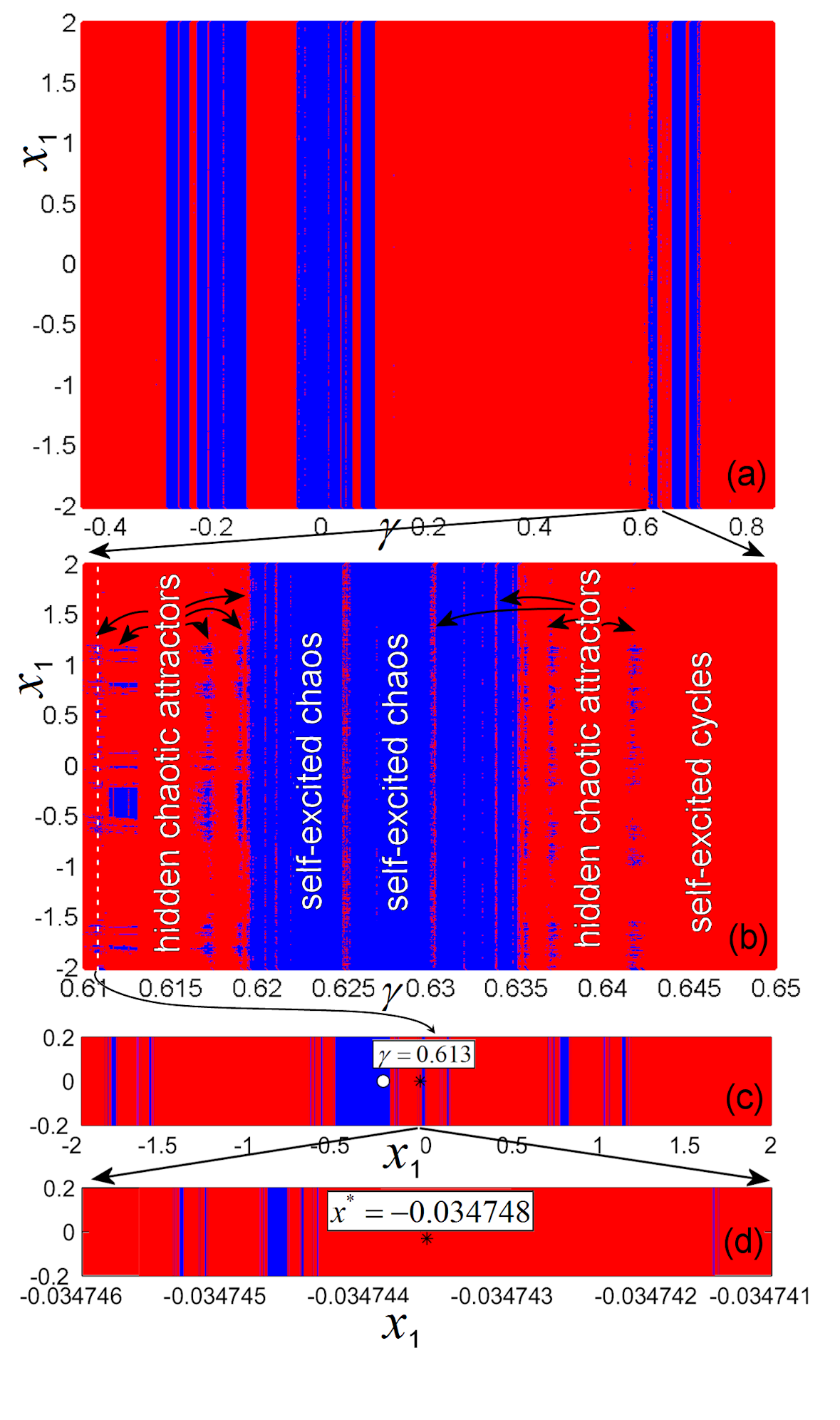}
\caption{Attraction basins of the impulsed SD system \eqref{e1d}-\eqref{supl} for $\De=1$ and $\ga\in[-0.45,0.85]$; (a) Contiguous regions representing the initial points which generate self-excited attractors (red for self-excited stable cycles and blue for self-excited chaotic attractors); (b) Zoomed image revealing supplementary interleaved, fractal-like regions, which represent the attraction basins of hidden chaotic attractors (blue) and self-excited stable cycles (red); (c) Attraction basin for a single value of $\ga$, $\ga=0.613$, with initial conditions $x_1\in[-2,2]$. The star represents the unstable equilibrium $x^*=-0.034748$, while the circle, $x_0=-0.25$; (d) Zoomed image around the unstable equilibrium $x^*$. For clarity, in Figures (c) and (d), the points $x_1$ are represented as vertical bars.}
\label{basinut}       
\end{center}
\end{figure*}

The attraction basins versus $\ga\in[-0.45,0.85]$ of the stable cycles and chaotic attractors, for $\De=1$, and $x_1\in[-2,2]$ as vertical axis, are presented in figure \ref{basinut} (a). The obtained parameter space, is partitioned in two kind of windows: stable windows (red plot), corresponding to stable cycles and chaotic windows (blue plot), corresponding to chaotic attractors. For whatever values $\ga$ in some of these windows, the impulsed system either evolves along a stable cycle or a chaotic attractor, for all initial conditions $x_1\in[-2,2]$. Therefore, the contiguous zone for $x_1\in[-2,2]$, represent the attraction basins of self-excited attractors (red zone for self-excited stable cycle and blue zone for self-excited chaotic attractors).

However, as can be seen in the detail around $\ga=0.6$, zoomed to $\ga\in[0.61,0.65]$ in figure \ref{basinut} (b), there are some ``hidden'' non-contiguous narrow windows, where the initial conditions $x_1$ are vertically disposed in fractal-like structure of subintervals.

In order to classify the two kind of coexisting attractors (stable cycles and chaotic attractors) within these particular zones, let consider $\ga=0.613$ within the parameter interval $\ga\in[0.61,0.65]$, corresponding to the dotted white line in figure \ref{basinut} (b) (see also the green line in figure \ref{fig5}). For this value of $\ga$, the corresponding attraction basin of initial points $x_1$ plotted for clarity as vertical bars, is presented in figure \ref{basinut} (c).
The interleaved bars recall a Cantor-like set: successive zooms (restricted to the computer precision) reveal new and new autosimilarity-like (figure \ref{basinut} (d)). This fractal structure of the attraction basins, underlines the fragility of these attractors.

\begin{figure*}
\begin{center}
\includegraphics[scale=0.54] {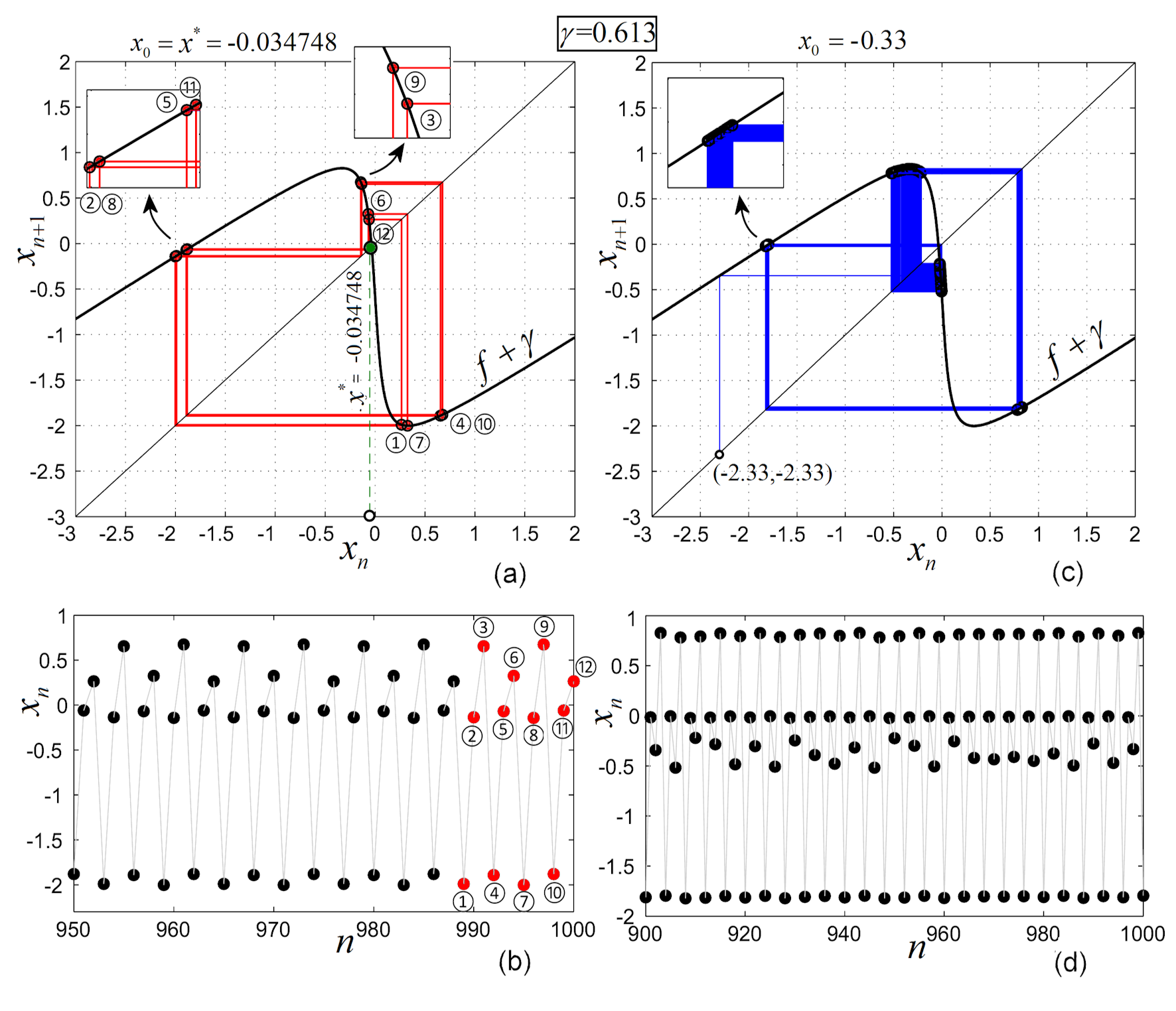}
\caption{Coexisting attractors for $\De=1$, $\ga=0.613$ and different initial conditions; (a), (b) Cobweb diagram and time series for $x_1=x^*=-0.034748$. Circled numbers represent the 11 stable elements of the cycle; (c), (d) Cobweb diagram and time series for $x_1=-0.33$.  }
\label{fig6}       
\end{center}
\end{figure*}

Let next consider the equilibrium point of the impulsed system, $x^*$, for $\De=1$ and $\ga=0.613$. To find $x^*$, one has to solve numerically the equation $F(x):=f(x)+\ga=x$ (see the relation \eqref{e2}), or to find the intersection of the graph $F(x)$ with the first bisector (figure \ref{fig6} (a)). With 6 decimals, $x^*=-0.034748$. The equilibrium is hyperbolic unstable: $|F'(x^*)|=|-13.454|>1$.

If one considers an initial condition in the red region, e.g. the point $x_1=x^*$, one obtains the stable cycle of period 5 (see the cobweb diagram in figure \ref{fig6} (a) and the time series in figure \ref{fig6} (b)). The order of fixed points which form this cycle is indicated by circled numbers. If one considers $x_1=-0.25$, for the same $\ga=0.613$ (circled blue point in figure \ref{basinut}) (c)) one obtains the chaotic attractor in figure \ref{fig6} (c), (d). Note that the initial value $x_1=x^*$ verifies Theorem \ref{teo} in this case too.

To conclude that the chaotic attractor is hidden, one has to verify that there exists some small neighborhood around $x^*$, wherefrom all initial conditions lead to the stable cycle and not to the hidden attractor. Numerically, the successive zooms of neighborhoods of the unstable fixed point $x^*$ (figure \ref{basinut} (d)), reveals the fact that there always exists a stable (red) neighborhood of points which lead to the stable cycle.

Moreover, the stable cycle is a self-excited attractor since it can be attained by initial conditions in the neighborhood of the unstable equilibrium.

Therefore, the stable cycle is self-excited attractor, while the chaotic attractor is hidden attractor.

\begin{figure*}
\begin{center}
\includegraphics[scale=0.65] {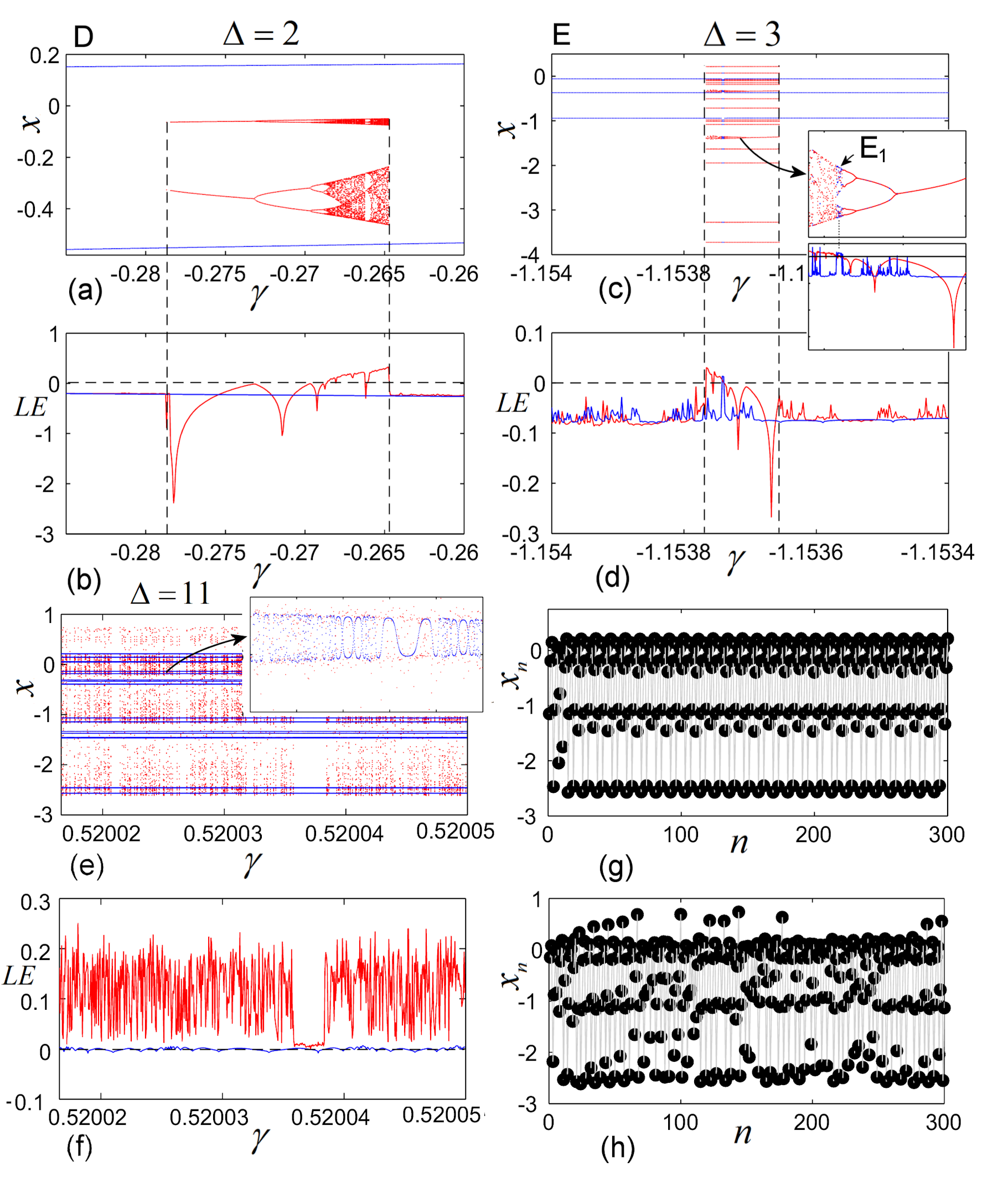}
\caption{Other hidden attractors: (a), (b) Overplotted bifurcation diagrams, LEs and zoomed regions, for $\De=2$ and $\ga\in[-0.29,-026]$; (c), (d) Overplotted bifurcation diagrams and zoomed region, and overplotted LEs for $\De=3$ and $\ga\in[-1.154,-1.1534]$; (e), (f) Bifurcation diagram and LEs for $\De=11$ and $\ga\in[0.52002,0.52005]$. Lyapunov exponent has an approximate zero value; (g) Time series of the self-excited mode-locking for $\ga=0.520037$ and $x_0=-0.04$; (h) Time series of the hidden chaotic attractor for $\ga=0.520037$ and $x_0=0.1$.}

\label{fig7}       
\end{center}
\end{figure*}

Hidden attractors are induced by the impulsive algorithm not only for $\De=1$, but also for all tested $\De$ values. For example, in figure \ref{fig7} are presented three hidden chaotic attractors and their self-excited stable cycles. The bifurcation diagrams and LE, reveal that similar with $\De=1$ (figure \ref{fig5}), for $\De=2$ (Figs \ref{fig7} (a), (b), zoomed detail $D$ from figure \ref{fig2} (a)) and $\De=3$ (Figs \ref{fig7} (c), (d), zoom detail $E$ from figure \ref{altii} (a)), the impulsed system presents hidden chaotic attractors which born and vanish suddenly, while the self-excited coexisting cycle still exist for the entire considered interval of $\ga$. Note that the zoomed detail $E_1$ reveals at, some value of $\ga$, an interchange between the hidden chaotic and self-excited stable cycle.

For $\De=11$, the hidden chaotic attractor has been found for a narrow interval of $\ga$ and, in contrast with the Feigenbaum route-like for $\De=1$, $\De=2$ and $\De=3$, presents intermittent-like irregularities (figure \ref{fig7} (e), (f)), and coexists with a multiple self-excited stable cycle (see the time series in figure \ref{fig7} (g), (h)). Note that the approximate zero value of the Lyapunov exponent in Fig. \ref{fig7} (f), and the time series (FIg. \ref{fig7} (g)) indicates a mode-locking orbit of the impulsed system for $\Delta=11$ and the chosen $\gamma$.

Compared to $\De=1$, where the hidden chaotic attractors appear following the Feigenbaum route, for $\De=3$ and also for $\De=11$, the system exhibits interior crisis, whereby the hidden chaotic attractors suddenly expand.

\section{Conclusion}
In this paper, it is shown that the chaotic dynamic of a discrete SD model can be controlled by constant impulses $\ga$ applied periodically every $\Delta$ steps. Also, it is analytically proved that the impulsed system admits periodic orbits and the obtained orbits are bounded for whatever impulse values and $\De$. The parameter $\ga$ can be either positive or negative, which suggests that the impulses can either increases or decreases its energy-like.
It is proved numerically and with aid of computer graphics that the impulses induces hidden chaotic attractors coexisting with self-excited stable cycles, fact which represents a novelty for discrete economic systems.
This capability of controlling chaotic systems can be useful in case of several other kind of systems such as systems interacting with the environment, as various life forms, chemical reactions, electrical systems, etc.

The impulsive algorithm can be used successfully to all discrete dynamical systems, but also to continuous or discontinuous (time-continuous) dynamical systems of integer or fractional order.

An open problem, which represents the subject of a future research, is to find conditions on $\Delta$ and $\ga$, to obtain targeted stable orbits.

\vspace{3mm}
\textbf{Acknowledgement }
M.-F. D. is supported by Tehnic B SRL and M. F. is supported by the Slovak Research and Development Agency under the contract No. APVV-14-0378 and by the Slovak Grant Agency VEGA No. 2/0153/16 and No. 1/0078/17.

\begin{thebibliography}{90}

\bibitem{eco}Lorenz H W, Nonlinear Dynamical Economics and Chaotic Motion, 2nd ed, Berlin: Springer-Verlag 1993
\bibitem{supli1} Beamon B M, Supply chain design and analysis: Models and methods International Journal of Production Economics, \textbf{55} (1998) 281-294
\bibitem{supli2}Harvey E B and Murthy K S R, Forecasting manpower demand and supply: A model for the accounting profession in Canada, International Journal of Forecasting, \textbf{4} (1988) 551-562
\bibitem{supli3} Sep\'{u}lveda-Rojas J P, Rojas F, Vald\'{e}s-González H and San Mart\'in M, Forecasting models selection mechanism for supply chain demand estimation Procedia Computer Science, \textbf{55} (2015) 1060-1068
    \bibitem{hid1} Leonov G A and Kuznetsov N V, Hidden attractors in dynamical systems. From hidden oscillations in Hilbert-Kolmogorov, Aizerman, and Kalman problems to hidden chaotic attractor in Chua circuits, International Journal of Bifurcation and Chaos, \textbf{23} (2013) 30002 69pp
\bibitem{hid2}Leonov G A, Kuznetsov N V and Mokaev T N, Homoclinic orbits, and self-excited and hidden attractors in a Lorenz-like system describing convective fluid motion European Physical Journal Special Topics, \textbf{224} (2015) 1421-1458
\bibitem{hid3} Kuznetsov  N V, Leonov G A, Mokaev T N, Prasad A and Shrimali M D, Finite-time Lyapunov dimension and hidden attractor of the Rabinovich system Nonlinear Dynamics \textbf{92} (2018) 267-285
\bibitem{jaf} Jafari S, Pham V-T, Golpayegani S M R H, Moghtadaei M and Kingni S T, The relationship between chaotic maps and some chaotic systems with hidden attractors, International Journal of Bifurcation and Chaos, \textbf{26} (2016) 1650211 8pp
\bibitem{jian1} Jiang H, Liu Y, Wei Z and Zhang L, A new class of three-dimensional maps with hidden chaotic dynamics, International Journal of Bifurcation and Chaos, \textbf{26} (2016) 1650206 13pp
\bibitem{jian2} Jiang H, Liu Y, Wei Z and Zhang L, Hidden chaotic attractors in a class of two-dimensional maps, Nonlinear Dynamics, \textbf{85} (2016) 2719-2727
\bibitem{nini} Andrievsky B R, Kuznetsov N V, Leonov G A and Pogromsky A Yu, Hidden oscillations in aircraft flight control system with input saturation, IFAC Proceedings Volumes, \textbf{46} (2013) 75-79
\bibitem{nini2}Lauvdal T, Murray R and Fossen T, Stabilization of integrator chains in the presence of magnitude and rate saturations: a gain scheduling approach. In Proc. IEEE Control and Decision Conference \textbf{4} (1997) 4404-4005
\bibitem{gm1}G\"{u}\'{e}mez J and Mat\`{\i}as M A, Control of chaos in unidimensional maps, Phys. Lett. A. \textbf{181} (1993) 29-32
\bibitem{gm2} Mat\`{\i}as M A and G\"{u}\'{e}mez J, Stabilization of chaos by proportional pulses in system variables Phys. Rev. Lett. \textbf{72} (1994) 1455-1458
\bibitem{multi1} Arecchi F T and Lisi F, Arecchi and Lisi respond, Phys. Rev. Lett. \textbf{50} (1983) 1330
\bibitem{multi2} Arecchi F T, Meucci R, Puccioni G and Tredicce J R, Experimental evidence of subharmonic bifurcations, multistability, and turbulence in a Q-switched gas laser, Phys. Rev. Lett. \textbf{49} (1982) 1217–1220
\bibitem{multi3} Saucedo-Solorio J M, Pisarchik A N, Kir'yanov A V and Aboites V, Generalized multistability in a fiber laser with modulated losses J. Opt. Soc. Am. B. \textbf{20} (2003) 490-496
\bibitem{multi4} Meucci R, Salvadori F, Naimee K A, Brugioni S, Goswami B K, Boccaletti S and Arecchi F T, Attractor selection in a modulated laser and in the Lorenz circuit, Phil. Trans. R. Soc. A, \textbf{366} (2008) 475-486
\bibitem{ex1} Danca M-F Chaos suppression via periodic change of variables in a class of discontinuous dynamical systems of fractional order, Nonlinear Dynam. \textbf{70} (2012) 815-823
\bibitem{ex2} Danca M-F Chaos suppression via periodic pulses in a class of piece-wise continuous systems, Comput. Math. App. \textbf{64} (2012) 849-855
\bibitem{ex3}Danca M-F, Tang W K S, Wang Q and Chen G, Suppressing chaos in fractional-order systems by periodic perturbations on system variables, Eur. Phys. J. B. \textbf{86} (2013) Art 79 8pp
\bibitem{ex4}Danca M-F, Tang W K S and Chen G, Suppressing chaos in a simplest autonomous memristor-based circuit of fractional order by periodic impulses, Chaos, Soliton. Fract. \textbf{84} (2016) 31-40
\bibitem{danca1}Danca M-F, Fe\v ckan M and Chen G, Impulsive stabilization of chaos in fractional-order systems, Nonlinear Dynam. \textbf{89} (2017) 1889-1903
\bibitem{dancax}Codreanu S and Danca M-F, Suppression of chaos in a one-dimensional mapping, J. Biol. Phys. \textbf{23} (1997) 1-9
\bibitem{dancay} Codreanu S and Danca M-F, Control of chaos in a nonlinear prey-predator model, Polish Journal of Environmental Studies, \textbf{6} (1997) 21-24
\bibitem{Z}Zhang W B, Discrete Dynamical Systems, Bifurcations, and Chaos in Economics, Boston: Elsevier, 2006
\bibitem{futo}Futoma H, Southworth B Discrete chaotic dynamical systems in economic models preprint (2015)
\bibitem{El} Elaydi S N, An Introduction to Difference Equations 3rd ed, New-York: Springer-Verlag, 2005
\bibitem{ei} Danca M-F, Fe\v{c}kan M and Posp\'{\i}\v{s}il M, Difference equations with impulses, Opuscula Math. \textbf{39}(1) (2019) 5-22





\end{thebibliography}
\end{document}